\documentclass[english,twocolumn]{revtex4-1}
\usepackage[T1]{fontenc}
\usepackage[latin9]{inputenc}
\usepackage{float}
\usepackage{mathrsfs}
\usepackage{bm}
\usepackage{multirow}
\usepackage{amsmath}
\usepackage{graphicx}

\makeatletter

\providecommand{\tabularnewline}{\\}

\usepackage{babel}

\makeatother

\usepackage{babel}
\begin{document}
\title{Realization of doubly inhomogeneous waveplates for structuring of
light beams}

\author{Radhakrishna B}
\thanks{These two authors contributed equally}

\author{Gururaj Kadiri}
\thanks{These two authors contributed equally}
\author{G. Raghavan}
\affiliation{Materials Physics Division, Materials Science Group, Indira Gandhi
Centre for Atomic Research, HBNI, Kalpakkam, 603102, India}
\email{Radhakrishna B: brkrishna@igcar.gov.in}

\begin{abstract}
Waveplates having spatially varying fast-axis orientation and retardance
provide an elegant and easy way to locally manipulate different attributes
of light beams namely, polarization, amplitude and phase, leading
to the generation of exotic structured light beams. The fabrication
of such doubly inhomogeneous waveplates (d-plates) is more complex,
compared to that of singly inhomogeneous waveplates (s-plates) having
uniform retardance, which can be easily fabricated by different means
such as photoalignment of liquid crystals, metasurfaces etc. Here,
exploiting the SU(2) formalism, we establish analytically that any
d-plate can be equivalently implemented using a pair of quarter-wave
s-plates and a half-wave s-plate. An important advantage of this method
is that it gives the flexibility to realize a whole family of distinct
d-plates using the same triplet of s-plates. To underline the scope
of this method, we propose novel d-plates for spatially tailoring
the phase and complex amplitudes of light beams. Towards complex amplitude
shaping, we present a generic method for carving out higher-order
eigenmodes of light using a d-plate in conjugation with a polarizer.
A generalized q-plate-like gadget, for imparting a polarization-dependent
phase profile to a scalar light beam, is proposed as a demonstration
of phase-polarization interplay. For these two illustrations, the
corresponding three-s-plate gadget is constructed, and its functioning
is validated with extensive numerical simulations. The main result
and its illustrations are generic and agnostic to the way the s-plates
are fabricated and we believe they carry the potential to push the
current state of the art in interdisciplinary applications involving
structured light beams. 
\end{abstract}
\maketitle

\section{Introduction}

Waveplates have been indispensable workhorses in optics laboratories,
extensively used for manipulating the state of polarization (SoP)
of light beams. Traditionally, they are made from optically anisotropic
materials and characterized by a uniform retardance and a unique orientation
of fast-axis confined to its plane, classified here as homogeneous
waveplates. One could also fabricate waveplates with spatial variation
in either retardance or fast-axis orientation or both, and are here
termed as inhomogeneous waveplates. For notational convenience, we
classify the waveplates based on their inhomogeneity, into four kinds,
as summarized in the Tab. (\ref{tab:Classification-of-waveretarders}).

\begin{table}[H]
\begin{centering}
\begin{tabular}{|c|c|c|c|}
\hline 
\multicolumn{2}{|c|}{Kind} & Retardance $\Gamma\left(r,\phi\right)$  & fast-axis $\alpha\left(r,\phi\right)$\tabularnewline
\hline 
\hline 
\multicolumn{2}{|c|}{Homogeneous} & Uniform  & Uniform\tabularnewline
\hline 
Singly inhomogeneous  & 1  & Uniform  & Nonuniform\tabularnewline
\cline{2-4} \cline{3-4} \cline{4-4} 
(s-plate)  & 2  & Nonuniform  & Uniform\tabularnewline
\hline 
\multicolumn{2}{|c|}{Doubly inhomogeneous} & \multirow{2}{*}{Nonuniform} & \multirow{2}{*}{Nonuniform}\tabularnewline
\multicolumn{2}{|c|}{(d-plate)} &  & \tabularnewline
\hline 
\end{tabular}
\par\end{centering}
\caption{Classification of waveplates in the order of increasing inhomogeneity.\label{tab:Classification-of-waveretarders}}
\end{table}

A light beam with a spatially uniform SoP in its transverse plane,
referred to as scalar beam, through a homogeneous waveplate exits
as a scalar beam with a different SoP. In this process, it also picks-up
a kind of geometric phase, called the Pancharatnam-Berry (PB) phase
\citep{pancharatnam1956generalized,berry1987adiabatic}, whose magnitude
depends on the retardance and fast-axis orientation of the waveplate,
in addition to the input SoP. On the other hand, a scalar beam through
an inhomogeneous waveplate would emerge out with a spatially varying
SoP and/or phase. Light beams with spatially varying SoPs are termed
as vector beams\citep{zhan2009cylindrical,qiwen2013vectorial}. Generation
of light beams with spatially varying phase, i.e., wavefront shaping,
is also possible using inhomogeneous waveplates, based on the idea
of PB-phase. For instance a well-known s-plate of the first kind is
the q-plate\citep{marrucci2006optical}, whose fast-axis varies linearly
with the azimuthal angle. The standard q-plates have a retardance
$\pi$ and a scalar light beam with circular polarization through
it acquires a helical wavefront, in addition to flipped helicity in
its polarization.

Such tailoring of light beams, with spatial variation in SoP, amplitude
and phase, leads to a bigger class of light beams called structured
light\citep{andrews2011structured,rubinsztein2016roadmap,rosales2018review}.
In the recent past, structured light has found increased applications.
For instance, radially polarized light, a kind of vector beams, are
shown to provide sharper focusing in comparison with linearly polarized
light \citep{dorn2003sharper}. Light beams with helical phase are
known to carry orbital angular momentum\citep{allen1992orbital,shen2019optical}
and this concept has given birth to many novel phenomena in optics
like: spin-orbit interactions\citep{bliokh2015spin}, spin-Hall effect\citep{liu2017photonic}
and these have shown promising applications too\citep{shitrit2013spin,aiello2015transverse}.
Structuring of light has also contributed in, among others, optical
trapping\citep{kozawa2010optical,roxworthy2010optical,taylor2015enhanced},
material processing\citep{meier2007material} and quantum information
tasks\citep{erhard2018twisted}.

The last couple of decades has therefore witnessed a flurry of activity
towards designing inhomogeneous waveplates aimed at the generation
of structured light beams. These efforts have been particularly successful
with respect to s-plates, leading to diverse methods of fabricating
them being established, prominent ones being those based on photoalignment
of liquid crystals\citep{Chigrinov2008,kim2015fabrication,ji2016meta}
and metasurfaces\citep{lin2014dielectric,khorasaninejad2016metalenses,scheuer2020optical}.
The richness offered by d-plates in applications is just beginning
to be explored\citep{ling2015giant,pal2016tunable,devlin2017arbitrary,rafayelyan2017laguerre,rubin2019matrix},
their adaptation being slow perhaps due to the challenges in their
fabrication. Use of metasurfaces for fabricating the d-plates is now
picking up steam\citep{arbabi2015dielectric,mueller2017metasurface},
but it involves precise control of dimensions in the nano-meter length
scale. Further, in this method the parameters of the d-plate get fixed
at the time of fabrication and cannot be dynamically tuned. In liquid
crystal based s-plates, on the other hand, it is possible to achieve
active control of retardance through external means like applied voltage
\citep{piccirillo2010photon,Slussarenko:11} or temperature\citep{Karimi2009},
but, to the best of our knowledge, there has been no literature on
fabrication of d-plates using liquid crystals.

These limitations in current methods of fabricating d-plates forces
one to seek alternate ways of realizing them. In this context, it
is interesting to explore whether a stack of s-plates can effectively
function like a d-plate. Such effective waveplates, have earlier assisted
in realizing many functionalities which otherwise are not possible
through individual waveplates. For instance, early work of Pancharatnam
demonstrated that a half-wave (HW) plate sandwiched between two identically
oriented quarter-wave (QW) plates functions effectively as a tunable
retardance waveplate and has been used for realizing achromatic waveplates\citep{pancharatnam_1_1955achromatic,pancharatnam2_1955achromatic}
and are even available commercially \citep{thorlabs_Pancharatnam_achromat}.
Stacking of s-plates has also yielded many novel results. For instance,
combination of q-plates and waveplates are used for changing the topological
charges of q-plates\citep{yi2015addition,delaney2017arithmetic}.
Passively tuning the retardance of q-plates is possible from a combination
of q-plates wherein, the retardance is tuned by merely changing the
relative orientation of the involved q-plates\citep{radhakrishna2019wavelength}.

In this article, we first demonstrate that an arrangement involving
a pair of identically oriented QW-s-plates with a HW-s-plate placed
in between (henceforth referred to as QHQ-s-plate) is effectively
equivalent to a d-plate. Conversely, given a d-plate, we establish
that there exists a unique QHQ-s-plate equivalent to it and derive
the fast-axis orientations of those s-plates. While local manipulation
of SoPs of light using waveplates is well-studied, spatial structuring
of its complex amplitude and phase is more involved and rarely discussed.
Therefore as a strong demonstration of this \textit{$\text{d-plate}\equiv\text{QHQ-s-plate}$}
equivalence, we theoretically propose novel d-plate based gadgets
for local manipulation of amplitude and phase of light beams. For
these two cases, we numerically simulate the corresponding QHQ-s-plate
and validate its ability to mimic the d-plate. These examples are
of significant interest and complexity in their own realm, and have
spawned a large amount of literature.

The first illustration is that of tailoring the complex amplitude
of light beams. Different varieties of light beams are studied in
the literature\citep{korotkova2013random}, for instance Laguerre-Gaussian(LG)
and Hermite-Gaussian(HG) beams\citep{galvez2017complex}, non-diffracting
beams like Bessel beams\citep{bouchal2003nondiffracting,mazilu2010light},
accelerating beams\citep{efremidis2019airy} and so on. Each of these
beams have found numerous applications, for instance, in reducing
the thermal noise of gravitational wave detectors \citep{granata2010higher,tao2020higher},
in STED microscopy\citep{hell1994breaking,yu2016super}, in optical
tweezers\citep{simpson1996optical,simpson1997mechanical} Bose-Einstein-condensation\citep{wright2000toroidal}
etc. Here, we demonstrate that starting from the fundamental Gaussian
mode of light, higher-order LG and HG beams can be generated using
a combination of d-plate and a polarizer. Similar strategy of realization
has been employed in \citep{rafayelyan2017laguerre,rafayelyan2017laguerre_nanostructure},
but restricted to higher order LG-beams.

The next illustration is meant at designing a d-plate that imparts
a polarization-dependent phase profile to the input scalar light beam.
This, in a sense, generalizes the notion of q-plate to arbitrary elliptical
polarization (instead of circular polarization) and arbitrary wavefront
(instead of helical wavefront). This example also serves to bring
out the limitation of d-plates in affecting PB-phase based SoP transformations.

The rest of the article is arranged as follows: Section \ref{sec:Theory}
presents theoretical analysis of our scheme for realizing d-plate,
and in section \ref{sec:Illustration} we provide two distinct case
studies as illustrations. We conclude the article by aggregating the
essential results in section \ref{sec:Summary-and-Conclusions}.

\section{\label{sec:Theory}Theory}

The SoP of a light beam refers to the direction of the time-varying
electric field vector in its transverse plane. SoPs are described
in different but equivalent ways, prominent ones being Jones vector,
Stokes vector and as points on the Poincare sphere. The Jones vector
$\left|\theta,\varphi\right\rangle $ of an SoP is a two-dimensional
complex vector of unit norm given by:

\begin{equation}
\left|\theta,\varphi\right\rangle ={\textstyle \cos\frac{\theta}{2}\vert L\rangle+e^{i\varphi}\sin\frac{\theta}{2}\vert R\rangle}\label{eq:Jones_Vector}
\end{equation}
where $\vert L\rangle$ and $\vert R\rangle$ are left and right circular
polarizations respectively, which in the horizontal-vertical basis,
$\vert H\rangle=\left(1,0\right)^{T}$and $\vert V\rangle=\left(0,1\right)^{T}$,
are chosen to be $\vert L\rangle=\frac{1}{\sqrt{2}}\left(\vert H\rangle+i\vert V\rangle\right)$
and $\vert R\rangle=\frac{1}{\sqrt{2}}\left(\vert H\rangle-i\vert V\rangle\right)$
and $0\leq\theta<\pi$, $0\leq\varphi<2\pi$. One could also characterize
the SoP $\left|\theta,\varphi\right\rangle $ by a three-dimensional
real unit vector $\bm{S}_{\theta,\varphi}$, called the Stokes vector
whose explicit expression is

\begin{equation}
\bm{S}_{\theta,\varphi}=\left[\sin\theta\cos\varphi,\:\sin\theta\sin\varphi,\:\cos\theta\right]^{T}\label{eq:Stokes_Vector}
\end{equation}
The Jones and Stokes vector representations of a polarization state
are connected by\citep{damask2004polarization}

\begin{equation}
\bm{S}_{\theta,\varphi}=\langle\theta,\varphi\vert\bm{\sigma}\vert\theta,\varphi\rangle
\end{equation}
where $\bm{\sigma}=\left(\sigma_{x},\sigma_{y},\sigma_{z}\right)$
is a Pauli-spin vector with components:

\begin{equation}
\sigma_{x}=\left[\begin{array}{cc}
1 & 0\\
0 & -1
\end{array}\right],\:\sigma_{y}=\left[\begin{array}{cc}
0 & 1\\
1 & 0
\end{array}\right],\:\text{and}\:\sigma_{z}=\left[\begin{array}{cc}
0 & -i\\
i & 0
\end{array}\right]
\end{equation}
These SoPs can also be represented geometrically as points on the
surface of a unit sphere, called the Poincare sphere. In this description,
the SoP $\left|\theta,\varphi\right\rangle $ is mapped to the point
whose polar and azimuthal coordinates are $\theta$ and $\varphi$
respectively.

SoP of a light beam can be altered, without changing its intensity,
by use of anisotropic optical elements called waveplates. They are
characterized by two parameters: retardance $\Gamma$ and orientation
of the fast-axis $\alpha$. They function by introducing a phase difference
of $\Gamma$ between the electric field component along the fast-axis
$\alpha$ and its orthogonal direction. For instance QW-plate and
HW-plate introduce a phase difference of $\Gamma=\frac{\pi}{2}$ and
$\pi$ respectively. As waveplates merely introduce phase difference
between the orthogonal components, they do not alter the intensity
of the light beam. Hence, their action is mathematically described
by norm-preserving matrices, which in Jones vector formalism are unitary
matrices called Jones matrices and in Stokes vector formalism are
orthogonal matrices.

By definition, Jones matrix of waveplates accounts for the linear
transformation between two Jones vectors of SoPs. Jones vectors, however,
are arbitrary up to a global phase factor, and hence waveplates that
transform the SoPs can be represented by unitary matrices, without
regard to the determinant. Nevertheless, for correct handling of the
associated PB phase change, it is preferable to represent Jones matrices
as unitary matrices of unit determinant, i.e., $SU(2)$ matrices \citep{damask2004polarization}.
Therefore, in the rest of the article, the Jones matrices of waveplates
are described using SU(2) matrices.

On the Poincare sphere, the action of a waveplate with parameters
$\Gamma$ and $\alpha$, on an SoP $\left|\theta,\varphi\right\rangle $
is described as a rotation of its Stokes vector $\bm{S}_{\theta,\varphi}$,
about the rotation axis $\left(\cos2\alpha,\sin2\alpha,0\right)$,
by the angle $\Gamma$. The Jones matrix of such a waveplate, denoted
as $\mathcal{W}_{\Gamma}\left(\alpha\right)$, in the $\left\{ \vert H\rangle,\vert V\rangle\right\} $
basis is given by:

\begin{widetext} 

\begin{align}
\mathcal{W}_{\Gamma}\left(\alpha\right) & ={\textstyle \cos\left(\frac{\Gamma}{2}\right)\bm{I}+i\sin\left(\frac{\Gamma}{2}\right)\left(\cos2\alpha\cdot\sigma_{x}+\sin2\alpha\cdot\sigma_{y}\right)}\label{eq:Waveplate_SU2}\\
 & =\left[\begin{array}{cc}
\cos\left(\frac{\Gamma}{2}\right)+i\sin\left(\frac{\Gamma}{2}\right)\cos2\alpha & i\sin\left(\frac{\Gamma}{2}\right)\sin2\alpha\\
i\sin\left(\frac{\Gamma}{2}\right)\sin2\alpha & \cos\left(\frac{\Gamma}{2}\right)-i\sin\left(\frac{\Gamma}{2}\right)\cos2\alpha
\end{array}\right]\in SU\left(2\right)\label{eq:SU2_Action}
\end{align}

\end{widetext} 

where $\bm{I}$ is the $2\times2$ identity matrix. It may be noted
that $\mathcal{W}_{\Gamma}\left(\alpha\right)$ is a symmetric matrix,
with diagonal elements being complex conjugates of each other, and
off-diagonal elements being purely imaginary.

For a stack of waveplates, their combined action is described by a
matrix obtained by multiplying the Jones matrix of each waveplates.
While the product of Jones matrices of waveplates is also an SU(2)
matrix, it need not be a Jones matrix of the form of Eq. (\ref{eq:SU2_Action}).
However, if the resulting matrix is also of this form, then the sequence
of waveplates effectively functions like a single waveplate $\mathcal{W}_{\Gamma_{e}}\left(\alpha_{e}\right)$,
having an effective retardance $\Gamma_{e}$ and an effective fast-axis
orientation $\alpha_{e}$ which can be determined from the SU(2) matrix:

\begin{align}
\Gamma_{e} & ={\textstyle 2\cos^{-1}\bigg(\text{trace}\left(\mathcal{W}_{\Gamma_{e}}\left(\alpha_{e}\right)\right)\bigg)},\label{eq:Gamma_E}\\
\alpha_{e} & =\frac{1}{2}\tan^{-1}\left(\frac{w_{12}}{w_{11}}\right)\label{eq:Alpha_E}
\end{align}
where $w_{jk}$ is the imaginary part of $\left[\mathcal{W}_{\Gamma_{e}}\left(\alpha_{e}\right)\right]_{jk}$.

As mentioned in the previous section, an arrangement of HW-plate sandwiched
between two identically oriented QW-plates acts like a single waveplate,
with effective retardance $\Gamma_{e}$ depending on the relative
orientation between the HW-plate and QW-plates, and effective fast-axis
orientation $\alpha_{e}$ inclined at an angle of $\frac{\pi}{4}$
with respect to the QW-plates. This arrangement of waveplates, called
here as QHQ-waveplate, enables realizing a tunable retardance waveplate
where the tuning is achieved by merely changing the relative orientation
of the plates.

The above Jones matrix formalism of waveplates can be extended to
s-plates and d-plates, except that now the matrix $\mathcal{W}_{\Gamma}\left(\alpha\right)$
is a function of the radial and azimuthal coordinates $\left(r,\phi\right)$
of the plate, through $\Gamma$ or $\alpha$ or both. For instance,
the standard q-plate\citep{marrucci2006optical} is the most studied
s-plate of the first kind, with retardance $\pi$, and fast-axis orientation
$\alpha$ varying linearly with the azimuthal angle as $\alpha\left(\phi\right)=q\phi+\alpha_{0}$,
where $q$ is the topological charge and $\alpha_{0}$ is the offset
angle. Its Jones matrix is therefore a function of azimuthal coordinate
$\phi$: 
\begin{equation}
\mathcal{W}_{\pi}\left(\phi\right)=i\left(\sigma_{x}\cos2\left(q\phi+\alpha_{0}\right)+\sigma_{y}\sin2\left(q\phi+\alpha_{0}\right)\right)
\end{equation}

Here we extend the notion of QHQ-waveplate to such inhomogeneous waveplates
and explore its effective behavior. Since the QHQ-arrangement is possible
only with s-plates of first kind, in the rest of the article, s-plate
always refers to the s-plate of first kind classified in the Tab.
(\ref{tab:Classification-of-waveretarders}), unless mentioned otherwise.

Consider a HW-s-plate with fast-axis orientation $\alpha_{H}\left(r,\phi\right)$
placed in between two identically oriented QW-s-plates having fast-axis
distribution $\alpha_{_{Q}}\left(r,\phi\right)$. We refer to this
QHQ arrangement of s-plates as the ``QHQ-s-plate''. The effective
retardance $\Gamma_{e}\left(r,\phi\right)$ and the effective fast-axis
orientation $\alpha_{e}\left(r,\phi\right)$ of the QHQ-s-plate is
(see appendix \ref{sec:Single-effective-waveplate.}):

\begin{align}
\Gamma_{e}\left(r,\phi\right) & =2\pi+4\left(\alpha_{Q}\left(r,\phi\right)-\alpha_{H}\left(r,\phi\right)\right),\label{eq:Panch-Gamma}\\
\alpha_{e}\left(r,\phi\right) & =\alpha_{_{Q}}\left(r,\phi\right)+\frac{\pi}{4}\label{eq:Pancha_Alpha}
\end{align}

\begin{figure}[h]
\begin{centering}
\includegraphics[scale=0.22]{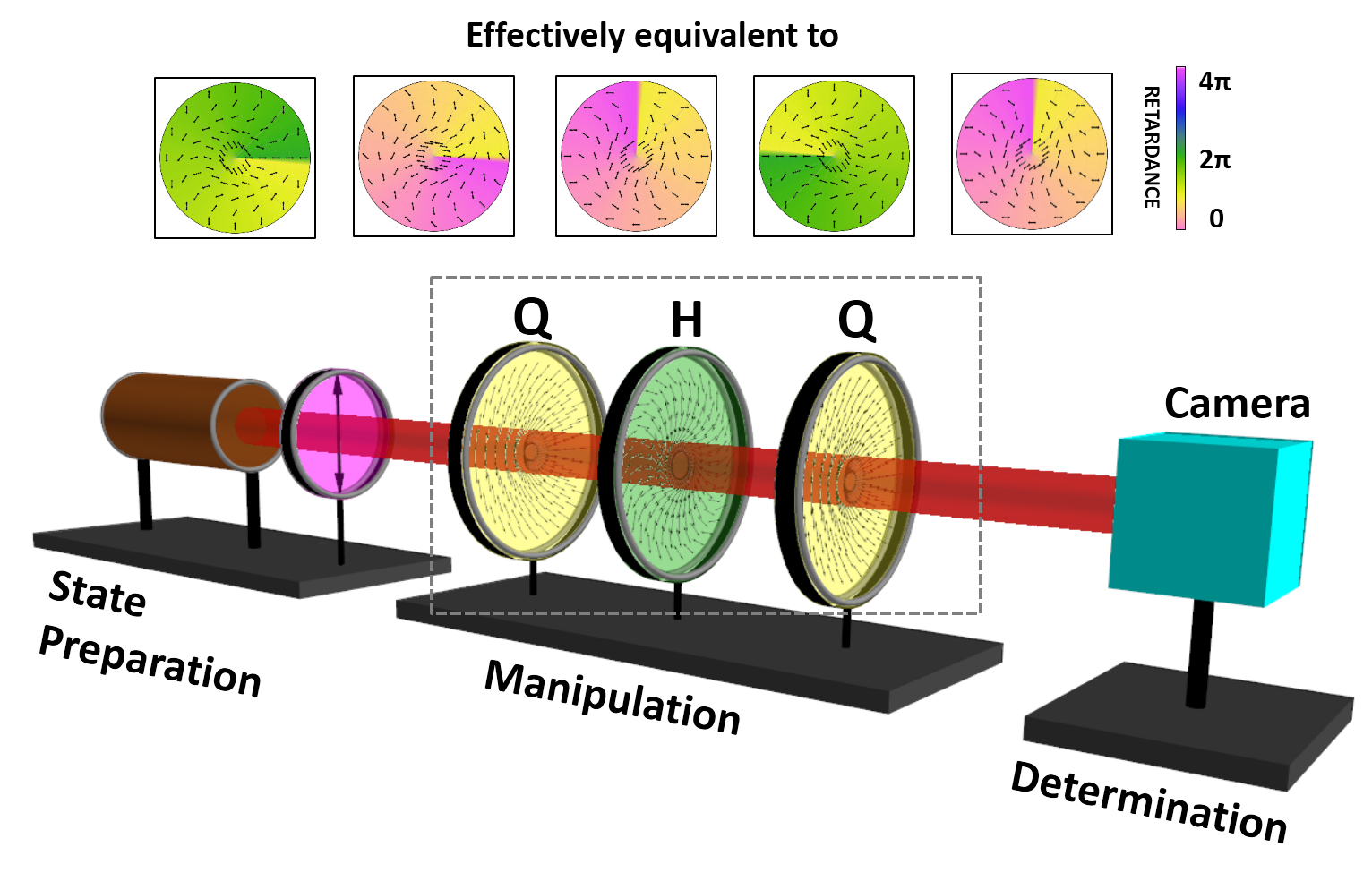} 
\par\end{centering}
\caption{Schematic for realizing effective d-plates using QHQ arrangement of
s-plates (color-coded by the retardance and arrows indicating the
fast-axis orientations). The different d-plates depicted on the top
correspond to different orientations of the same set of QHQ s-plates.\label{fig:Schematic_QHQ}}
\end{figure}

It should be noted that the effective retardance $\Gamma_{e}$ acquires
spatial dependence, and thereby converts the three s-plates into an
effective d-plate. Conversely, and more importantly, any desired d-plate
with $\Gamma_{e}\left(r,\phi\right)$ and $\alpha_{e}\left(r,\phi\right)$
can be uniquely realized as QHQ-s-plate with $\alpha_{_{Q}}\left(r,\phi\right)$
and $\alpha_{H}\left(r,\phi\right)$ given by:

\begin{align}
\alpha_{_{Q}}\left(r,\phi\right)=\, & \alpha_{e}\left(r,\phi\right)-\frac{\pi}{4},\label{eq:alpha_Q}\\
{\textstyle \alpha_{H}\left(r,\phi\right)=\,} & {\textstyle \alpha_{e}\left(r,\phi\right)+\frac{1}{4}\left(\pi-\Gamma_{e}\left(r,\phi\right)\right)}\label{eq:alpha_H}
\end{align}

It should be emphasized that the QHQ-s-plate is completely equivalent
to the d-plate, not just in bringing about the SoP transformations
but also in capturing the associated PB phases correctly. Moreover,
an important advantage of realizing d-plate through this method is
that a single set of QHQ s-plates can be employed for generating a
variety of d-plates. This is possible by merely changing the orientations
of three s-plates, constraining to the QHQ arrangement. 

To elaborate, consider a QHQ-s-plate, having fast-axis variation of
QW-s-plate and HW-s-plate as $\alpha_{Q}\left(r,\phi\right)$ and
$\alpha_{H}\left(r,\phi\right)$ respectively, set for realizing a
d-plate with parameters $\Gamma_{e}\left(r,\phi\right)$ and $\alpha_{e}\left(r,\phi\right)$.
Rotating these s-plates by angles $\delta_{Q}$ and $\delta_{H}$
respectively about their axes results in new s-plates having $\alpha_{Q}^{new}\left(r,\phi\right)$
and $\alpha_{H}^{new}\left(r,\phi\right)$, see Eq. (\ref{eq:Rotated_Fast_Axis_Orientation})
of appendix (\ref{sec:Rotating-the-waveplate}). These new s-plates
continue to remain in the QHQ arrangement and therefore yeild a new
d-plate, with parameters $\Gamma_{e}^{new}\left(r,\phi\right)$ and
$\alpha_{e}^{new}\left(r,\phi\right)$ given by:

\begin{widetext}

\begin{eqnarray}
\Gamma_{e}^{new}\left(r,\phi\right) & = & 2\pi+4\left(\alpha_{Q}\left(r,\phi-\delta_{Q}\right)-\alpha_{H}\left(r,\phi-\delta_{H}\right)+\left(\delta_{Q}-\delta_{H}\right)\right)\label{eq:Gamma_e_new}\\
\alpha_{e}^{new}\left(r,\phi\right) & = & \alpha_{Q}\left(r,\phi-\delta_{Q}\right)+\delta_{Q}+{\textstyle \frac{\pi}{4}}\label{eq:Alpha_e_new}
\end{eqnarray}

\end{widetext}

The five insets in Fig. (\ref{fig:Schematic_QHQ}) depict the distinct
d-plates realized using the same physical set of three s-plates but
by rotating them constraining to QHQ arrangement. 

In the following section, we explore the possible applications of
these results towards spatial control of amplitude and phase of light
beams.

\section{\label{sec:Illustration}illustrations of structuring the light beam
using d-plates}

\subsection{Tailoring the complex amplitude of light beams}

In this subsection, a generic method for spatial tailoring of the
amplitude and phase of light beams, using d-plates is proposed. As
a demonstration of this technique, realization of higher order LG
beams and HG beams starting from a fundamental Gaussian beam is discussed.
We aim to tailor a desired complex electric field $E_{des}\left(r,\phi\right)$,
starting from a scalar light beam having electric field $E_{in}\left(r,\phi\right)$.
We achieve this by transforming the input scalar light beam using
a d-plate, such that one of the components of the emerging vector
beam, in a particular orthogonal basis, is $E_{des}\left(r,\phi\right)$.
This component can be extracted by projecting out the orthogonal component.
For concreteness, we work in $\left\{ \vert H\rangle,\vert V\rangle\right\} $
basis, and let the SoP of the initial light beam $E_{in}\left(r,\phi\right)$
be $\vert H\rangle$. The transformation briefed above can be achieved
by an $SU\left(2\right)$ transformation $\mathcal{T}\left(r,\phi\right)$:

\begin{align}
\mathcal{T}\left(r,\phi\right)\,\left(E_{in}\left(r,\phi\right)\vert H\rangle\right) & =E_{des}\left(r,\phi\right)\vert H\rangle+E_{rem}\left(r,\phi\right)\vert V\rangle
\end{align}
where $E_{rem}\left(r,\phi\right)$ is the remnant electric field,
satisfying the relation: 
\begin{equation}
\vert E_{des}\vert^{2}+\vert E_{rem}\vert^{2}=\vert E_{in}\vert^{2}\label{eq:Conversation_Condition}
\end{equation}

Assuming $E_{in}\left(r,\phi\right)$ does not vanish within the region
of interest,

\begin{align}
\mathcal{T}\left(r,\phi\right)\,\vert H\rangle & =\frac{E_{des}\left(r,\phi\right)}{E_{in}\left(r,\phi\right)}\vert H\rangle+\frac{E_{rem}\left(r,\phi\right)}{E_{in}\left(r,\phi\right)}\vert V\rangle\label{eq:non_vanishing}
\end{align}

Given the action of $\mathcal{T}\left(r,\phi\right)$ on $\vert H\rangle,$
its action on vertical polarization $\vert V\rangle$ gets fixed because
of its $SU\left(2\right)$ property:

\begin{align}
\mathcal{T}\left(r,\phi\right)\,\vert V\rangle & =-\left(\frac{E_{rem}\left(r,\phi\right)}{E_{in}\left(r,\phi\right)}\right)^{*}\vert H\rangle+\left(\frac{E_{des}\left(r,\phi\right)}{E_{in}\left(r,\phi\right)}\right)^{*}\vert V\rangle
\end{align}
where $*$ denotes the complex conjugation. Hence, the matrix of $\mathcal{T}\left(r,\phi\right)$
in the $\left\{ \vert H\rangle,\vert V\rangle\right\} $ basis is
given by

\begin{equation}
\mathcal{T}\left(r,\phi\right)=\left[\begin{array}{cc}
\frac{E_{des}}{E_{in}}\quad & -\left(\frac{E_{rem}}{E_{in}}\right)^{*}\\
\\
\frac{E_{rem}}{E_{in}}\quad & \left(\frac{E_{des}}{E_{in}}\right)^{*}
\end{array}\right]
\end{equation}

The $SU(2)$ matrix $\mathcal{T}\left(r,\phi\right)$ will correspond
to the Jones matrix of a d-plate, as in Eq. (\ref{eq:SU2_Action}),
provided its off-diagonal elements are purely imaginary. The ratio
appearing along the off-diagonal is $\frac{E_{rem}}{E_{in}}=\frac{\vert E_{rem}\vert}{\vert E_{in}\vert}e^{i\left(\delta_{rem}-\delta_{in}\right)}$,
where $\delta_{rem}$ and $\delta_{in}$ are the respective phases
which, in general, can be spatially variant. As only the magnitude
of $E_{rem}\left(r,\phi\right)$ is fixed through Eq. (\ref{eq:Conversation_Condition})
leaving us free with the choice of $\delta_{rem}$, the off-diagonal
entries can be rendered purely imaginary by setting $\delta_{rem}=\delta_{in}+\frac{\pi}{2}$,
so that the resulting matrix can then be realized by a d-plate $\mathcal{W}_{\Gamma}\left(\alpha\right)$.
Placing a polarizing beam splitter at the state determination stage
(see Fig. (\ref{fig:Schematic_QHQ})), yields the desired field in
the horizontal arm and the remnant field in the vertical arm.

The required retardance $\Gamma\left(r,\phi\right)$ and fast-axis
orientation $\alpha\left(r,\phi\right)$ of the d-plate are extracted
from the resulting matrix through Eqs. (\ref{eq:Gamma_E} and \ref{eq:Alpha_E}):

\begin{eqnarray}
\cos{\textstyle \frac{\Gamma\left(r,\phi\right)}{2}} & {\textstyle =} & {\textstyle \frac{\left|E_{des}\right|}{\left|E_{in}\right|}\cos\left(\delta_{des}-\delta_{in}\right),}\label{eq:modal_gamma}\\
\tan2\alpha{\textstyle \left(r,\phi\right)} & = & {\textstyle \frac{\left|E_{rem}\right|}{\left|E_{des}\right|}\,\frac{1}{\sin\left(\delta_{des}-\delta_{in}\right)}}\label{eq:modal_alpha}
\end{eqnarray}

It follows from Eqs. (\ref{eq:modal_gamma} and \ref{eq:modal_alpha})
that for the case of $\delta_{des}=\delta_{in}$, the fast-axis orientation
$\alpha{\textstyle \left(r,\phi\right)}$ equals $\frac{\pi}{4}$
at all $\left(r,\phi\right)$, reducing the d-plate to an s-plate
of the second kind (see Tab. \ref{tab:Classification-of-waveretarders}).

Here, we apply this idea towards simulating LG and HG modes of higher
order. The functional forms of these modes are readily available,
for instance\citep{galvez2017complex}. At the location of the beam
waist $\left(z=0\right)$, they simplify to:

\begin{equation}
E_{l,p}\left(r,\phi;\mathscr{A},w\right)=\frac{\mathscr{A}}{w}\left(\frac{\sqrt{2}r}{w}\right)^{\vert l\vert}\,\mathcal{L}_{p}^{\vert l\vert}\left(\frac{2r^{2}}{w^{2}}\right)\,e^{-\frac{r^{2}}{w^{2}}}\,e^{il\phi}\label{eq:higher-order-LG-beam}
\end{equation}

\begin{equation}
E_{m,n}\left(x,y;\mathscr{A},w\right)=\frac{\mathscr{A}}{w}\mathcal{H}_{m}\left(\frac{\sqrt{2}x}{w}\right)\,\mathcal{H}_{n}\left(\frac{\sqrt{2}y}{w}\right)\,e^{-\frac{x^{2}+y^{2}}{w^{2}}}\label{eq:higher-order-HG-beam}
\end{equation}

where $\mathcal{L}_{p}^{\vert l\vert}$ denotes the associated Laguerre
polynomials with $l$ and $p$ being the azimuthal and radial indices
of the LG beam respectively; $\mathcal{H}_{m}$ refers to the Hermite
polynomial of order $m$; $w$ is the beam waist and $\mathscr{A}$
is a constant, indicating the power of the light beam.

We chose the input beam as the horizontally polarized fundamental
Gaussian mode, since it is non-vanishing at all finite values of $\left(r,\phi\right)$,
thereby satisfying the condition required for Eq. (\ref{eq:non_vanishing}).
Placing the d-plate at the location of the input beam waist, we have

\begin{equation}
E_{in}=E_{0,0}\left(r,\phi;\mathscr{A}_{in},w_{in}\right)\label{eq:modal_Ein}
\end{equation}

The LG and HG modes simulated are $E_{des}=E_{l,p}\left(r,\phi;\mathscr{A}_{des},w_{des}\right)$
and $E_{des}=E_{m,n}\left(x,y;\mathscr{A}_{des},w_{des}\right)$ respectively.
The possible values for $\mathscr{A}_{des}$ and $w_{des}$ is restricted
through Eq. (\ref{eq:Conversation_Condition}) which demands $\vert E_{des}\left(r,\phi\right)\vert\leq\vert E_{in}\left(r,\phi\right)\vert$
at all $\left(r,\phi\right)$ within the region of interest in the
transverse plane, which in turn dictate the conversion efficiency
of the d-plates.

\begin{figure}[h]
\begin{centering}
\includegraphics[scale=0.22]{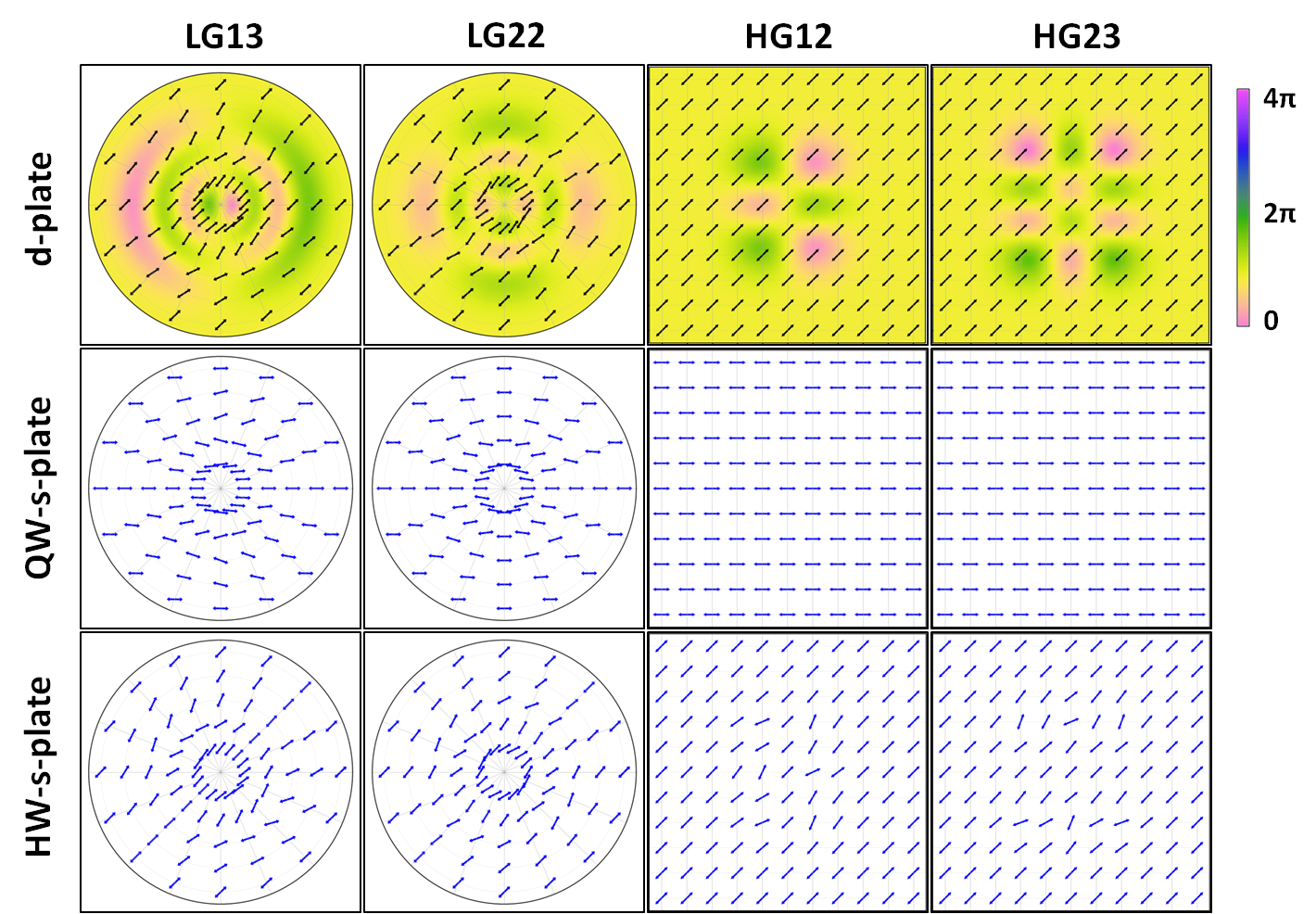} 
\par\end{centering}
\caption{Prototype of d-plates and the corresponding s-plates required for
realizing the higher order LG and HG modes starting from the fundamental
Gaussian mode. The top row depicts the d-plate whose spatial distribution
of retardance is coded in color and its fast-axis orientations depicted
by arrows. Second and third rows depict the fast-axis orientations
of the QW-s-plates and HW-s-plate respectively required for realizing
the above d-plates. The first and second column correspond to LG beam
with $\left(l,p\right)=\left(1,2\right)$ and $\left(2,3\right)$
respectively, while the third and fourth columns correspond to HG
beam with $\left(m,n\right)=\left(1,2\right)$ and $\left(2,3\right)$
respectively. \label{fig:modal_plates}}
\end{figure}

We now illustrate the working of this method by studying a few cases
through numerical simulations. We consider $\left(l,p\right)$=$\left(1,2\right)$
and $\left(2,3\right)$ in case of LG modes, and in case of HG modes,
$\left(m,n\right)=\left(1,2\right)$ and $\left(2,3\right)$. The
values of $\mathscr{A}_{des}$ and $w_{des}$ in all four cases are
taken to be $\frac{2\mathscr{A}_{in}}{3}$ and $\frac{w_{in}}{3}$
respectively. The retardance and fast-axis orientations of the d-plates
required for this apodization are calculated using Eqs. (\ref{eq:modal_gamma}
and \ref{eq:modal_alpha}). The required fast-axis orientations of
the QW-s-plates and HW-s-plate for realizing these d-plates are respectively
determined using Eqs. (\ref{eq:alpha_Q} and \ref{eq:alpha_H}) and
are summarized in Fig. (\ref{fig:modal_plates}).

\begin{figure}[h]
\begin{centering}
\includegraphics[scale=0.25]{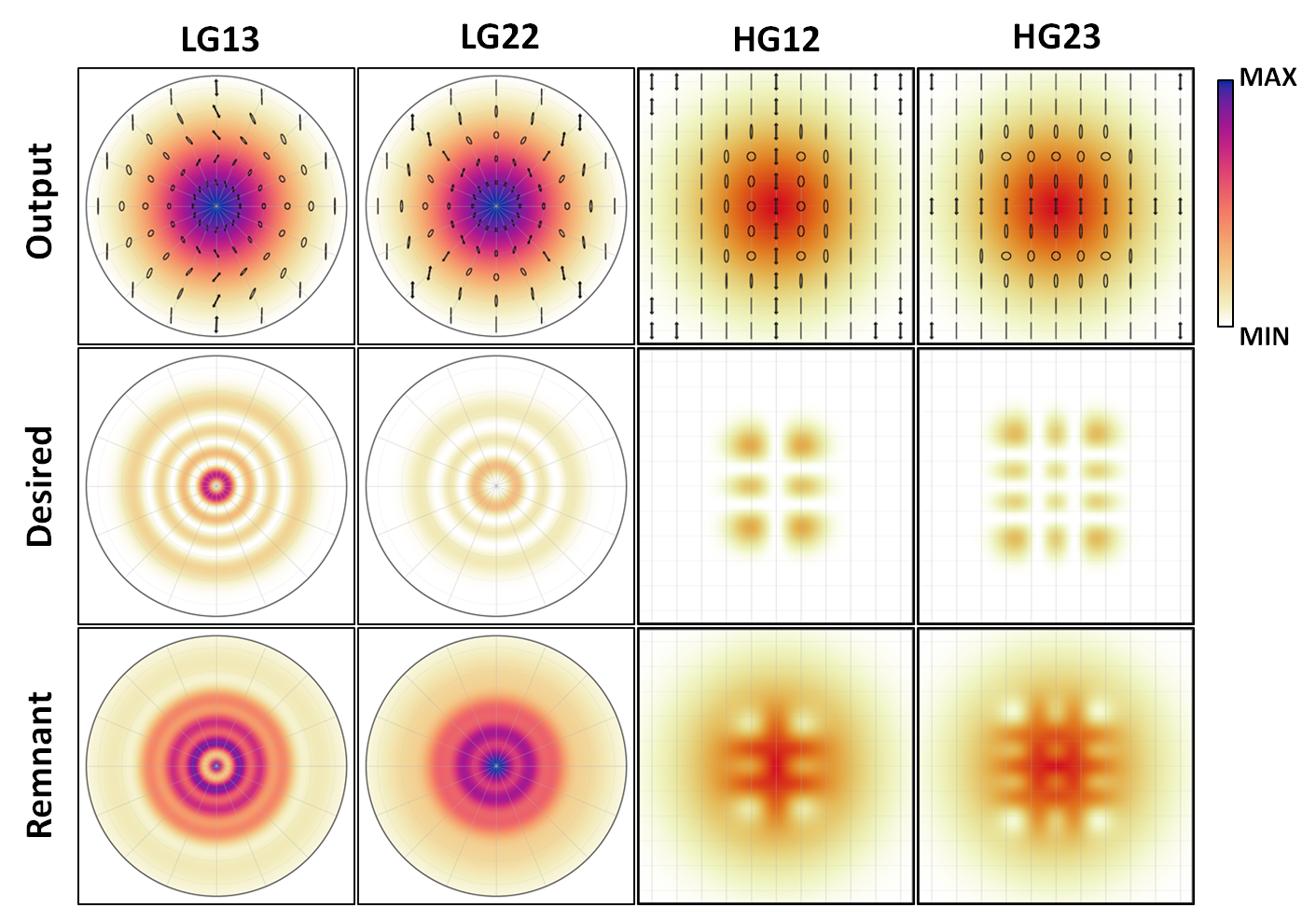} 
\par\end{centering}
\caption{Intensity profile of higher order LG and HG beams simulated numerically
using the QHQ arrangement of s-plates and polarizing beam splitters.
Along with intensity, the top row depicts SoPs distribution at the
exit plane of the QHQ arrangement of s-plates. Second and third rows
are the intensities projected along horizontal and vertical directions
corresponding to the desired $\left(\left|E_{des}\right|^{2}\right)$
and remnant $\left(\left|E_{rem}\right|^{2}\right)$ intensities respectively.
\label{fig:Modal_Intensity_Projection}}
\end{figure}

To validate the conversion to higher order modes, evolution of light
through the QHQ-s-plates is numerically simulated, ignoring the diffraction
and propagation effects. The input scalar beam, as it emerges out
of the QHQ-s-plate, gets converted into a vector beam through Eq.
(\ref{eq:SU2_Action}). The numerically simulated intensity and SoP
distributions in the transverse plane at the exit plane of these QHQ-s-plates
is depicted in Fig. (\ref{fig:Modal_Intensity_Projection}). The desired
modes and remnant fields are contained in the horizontal and vertical
component of these vector beams, which can be extracted by projecting
them, for instance, using a polarizing beam splitter. Figure (\ref{fig:Modal_Intensity_Projection})
summarizes the intensity distributions, $\left|E_{des}\right|^{2}$
and $\left|E_{res}\right|^{2}$ for all the four cases. The spatial
distribution of intensity in the horizontal projection is consistent
with that of the desired modes. The ratio of beam waists of output
and input beams appears to be a crucial parameter in deciding the
efficiency of conversion, as also observed in \citep{rafayelyan2017laguerre}.

To summarize this subsection, d-plate, together with a standard polarizer,
can aid in sculpting any complex amplitude light beams, out of the
ubiquitous fundamental Gaussian beams.

\subsection{Arbitrary polarization-dependent wavefront shaping }

Waveplates, apart from manipulating the SoP, also impart a total phase
(=dynamic phase + geometric phase), which depends on the plate parameters
and the involved SoPs. Given a pair of SoPs, there always exists a
waveplate that transforms one state to other unitarily and picking
up a particular phase in the process, depending on the path of transformation.
Owing to the spatial inhomogeneity of their fast-axis orientation,
s-plates are capable of imparting desired phase distribution onto
the input light beam. In these lines, the most studied s-plates are
the q-plates which basically flip the handedness of the input circular
polarization, and impart a helical phase in the process. In this subsection,
we seek to extend this functioning of q-plates to arbitrary SoPs instead
of circular polarizations, and to arbitrary phase distributions instead
of helical phase. The mathematical treatment required in the conversion
of an arbitrary SoP into an SoP of flipped handedness, and with arbitrary
phase distribution is first discussed. A d-plate based gadget is designed
to impart a polarization-dependent phase distribution to the input
light beam.

In general, using a waveplate it is not possible to transform SoPs
with a given phase, as the possible rotation axes for transforming
them are limited to two directions, thereby restricting the number
of phases to two. This stems from the fact that, the midplane of the
SoPs involved intersects the equatorial plane at only two points on
the Poincare sphere (which are antipodal), unless the midplane coincides
with the equatorial plane (see Fig. 1 in the ref. \citep{Bettegowda2017}).
Such a case occurs when the SoPs involved are enantiogyres\citep{salazar2018trajectories},
that is, mirror reflections of each other about the equatorial plane
of the Poincare sphere. Since the midplane of the enantiogyres coincides
with the equatorial plane of the Poincare sphere, every fast-axis
orientation $\alpha$ of the waveplate is a rotation axis for transforming
the states. This enables accessing every possible path in unitary
transformation of enantiogyres\citep{devlin2017arbitrary}. In other
words, for a waveplate oriented at any angle $\alpha\in\left(0,\pi\right)$,
there always exists a retardance $\Gamma$ depending on $\alpha$,
such that $\mathcal{W}_{\Gamma}\left(\alpha\right)$ takes an SoP
to its enantiogyre. Each one of these $\mathcal{W}_{\Gamma}\left(\alpha\right)$
generate a different phase between $0$ to $2\pi$ in the process.

Conversely, given a pair of enantiogyres and a desired spatial phase
distribution, one could always conceive of a d-plate $\mathcal{W}_{\Gamma}\left(\alpha\right)$
affecting such a transformation. In this subsection, we provide a
general prescription for constructing these d-plates, and their realization
using QHQ arrangement of s-plates. This exercise demonstrates, tailoring
the wavefront of a light beam and also assist to achieve an arbitrary
spin-orbit conversion.

Towards this, we seek a unitary transformation $\mathcal{U}$, that
transforms the SoP $\left|\theta,\varphi\right\rangle $ to its enantiogyre
$\left|\pi-\theta,\varphi\right\rangle $:

\begin{eqnarray}
\mathcal{U}\left|\theta,\varphi\right\rangle  & = & e^{i\psi\left(r,\phi\right)}\left|\pi-\theta,\varphi\right\rangle \label{eq:U_Transformation}
\end{eqnarray}

Being a unitary transformation, $\mathcal{U}$ naturally transforms
$\left|\pi-\theta,\pi+\varphi\right\rangle $ (the orthogonal state
of $\left|\theta,\varphi\right\rangle $) to $\left|\theta,\pi+\varphi\right\rangle $
(the orthogonal state of $\left|\pi-\theta,\varphi\right\rangle $):

\begin{equation}
\mathcal{U}\left|\pi-\theta,\pi+\varphi\right\rangle =e^{i\chi\left(r,\phi\right)}\left|\theta,\pi+\varphi\right\rangle \label{eq:U_Transformation_Antipodal}
\end{equation}
where $\psi\left(r,\phi\right)$ and $\chi\left(r,\phi\right)$ are
arbitrary spatially varying phases. The determinant of this unitary
transformation is $e^{i\left(\psi+\chi\right)}$ (see appendix \ref{ap:J_plate}).
Since, waveplates are mathematically SU(2) (determinant=$1$), they
alone cannot bring about this transformation. However any unitary
transformation $U(2)$ is equivalent to $U(1)\cdot SU(2)$\citep{simon1989universal,bhandari1997polarization},
hence the above transformation is possible by employing a phase plate
$\left(U\left(1\right)\right)$ for inducing a phase of $\left(\frac{\psi+\chi}{2}\right)$,
in conjugation with a waveplate:

\begin{equation}
\mathcal{U}=e^{i\left(\frac{\psi+\chi}{2}\right)}\boldsymbol{I}\cdot\mathcal{W}_{\Gamma}\left(\alpha\right)
\end{equation}
where $e^{i\left(\frac{\psi+\chi}{2}\right)}\boldsymbol{I}$ indicates
the action of phase plate. The required retardance $\Gamma$ and fast-axis
orientation $\alpha$ of the waveplate depend both on the SoP parameters
$\theta$,$\varphi$ and the difference of the phases $\psi$,$\chi$
(see appendix \ref{ap:J_plate}):

\begin{align}
{\textstyle \cos\frac{\Gamma}{2}=} & \sin\theta\cos\Delta\label{eq:J_Plate_Gamma}\\
\tan2\alpha= & \frac{\sin\left(\Delta-\eta\right)\sqrt{\sin^{2}\varphi+\cos^{2}\varphi\cos^{2}\theta}}{\sin^{2}\frac{\theta}{2}\sin\left(\Delta-\varphi\right)+\cos^{2}\frac{\theta}{2}\sin\left(\Delta+\varphi\right)}\label{eq:J_Plate_Alpha}
\end{align}
where $\Delta\left(r,\phi\right)=\frac{\psi-\chi}{2}$ and $\eta=\tan^{-1}\left(\cot\varphi\cos\theta\right)$.

For transformation between special enantiogyres left circular polarization
$\left(\left|0,0\right\rangle \right)$ and right circular polarization
$\left(\left|\pi,0\right\rangle \right)$, the retardance is uniformly
$\pi$, independent of the phases, and the fast-axis orientation is
of the form $\alpha\left(r,\phi\right)=\frac{\pi}{4}+\frac{\Delta}{2}$,
indicating that the necessary waveplate is a HW-s-plate. For any other
enantiogyres, parameters $\Gamma$ and $\alpha$ both acquire a spatial
dependence through $\Delta$, thereby demanding $\mathcal{W}_{\Gamma}\left(\alpha\right)$
to be a d-plate. As d-plate can be realized by the QHQ arrangement
of s-plates, augmenting this arrangement with a phase plate makes
it possible to generate scalar beams having any desired spatial variation
of phase.

As a concrete illustration of this idea, we now study the so-called
J-plate\citep{mueller2017metasurface,devlin2017arbitrary} proposed
for converting an arbitrary SAM to OAM states of light. J-plate extends
the notion of q-plates for circularly polarized light to arbitrary
elliptically polarized light. For J-plates, the phase distribution
of Eqs. (\ref{eq:U_Transformation} and \ref{eq:U_Transformation_Antipodal})
are of the form $\psi\left(r,\phi\right)=m\phi$ and $\chi\left(r,\phi\right)=n\phi$,
where $m$ and $n$ are integers. The phase plate required for this
turns out to be a spiral phase plate of order $\frac{m+n}{2}$\citep{beijersbergen1994helical}.

\begin{figure}[h]
\centering{}\centering{}\includegraphics[width=\linewidth]{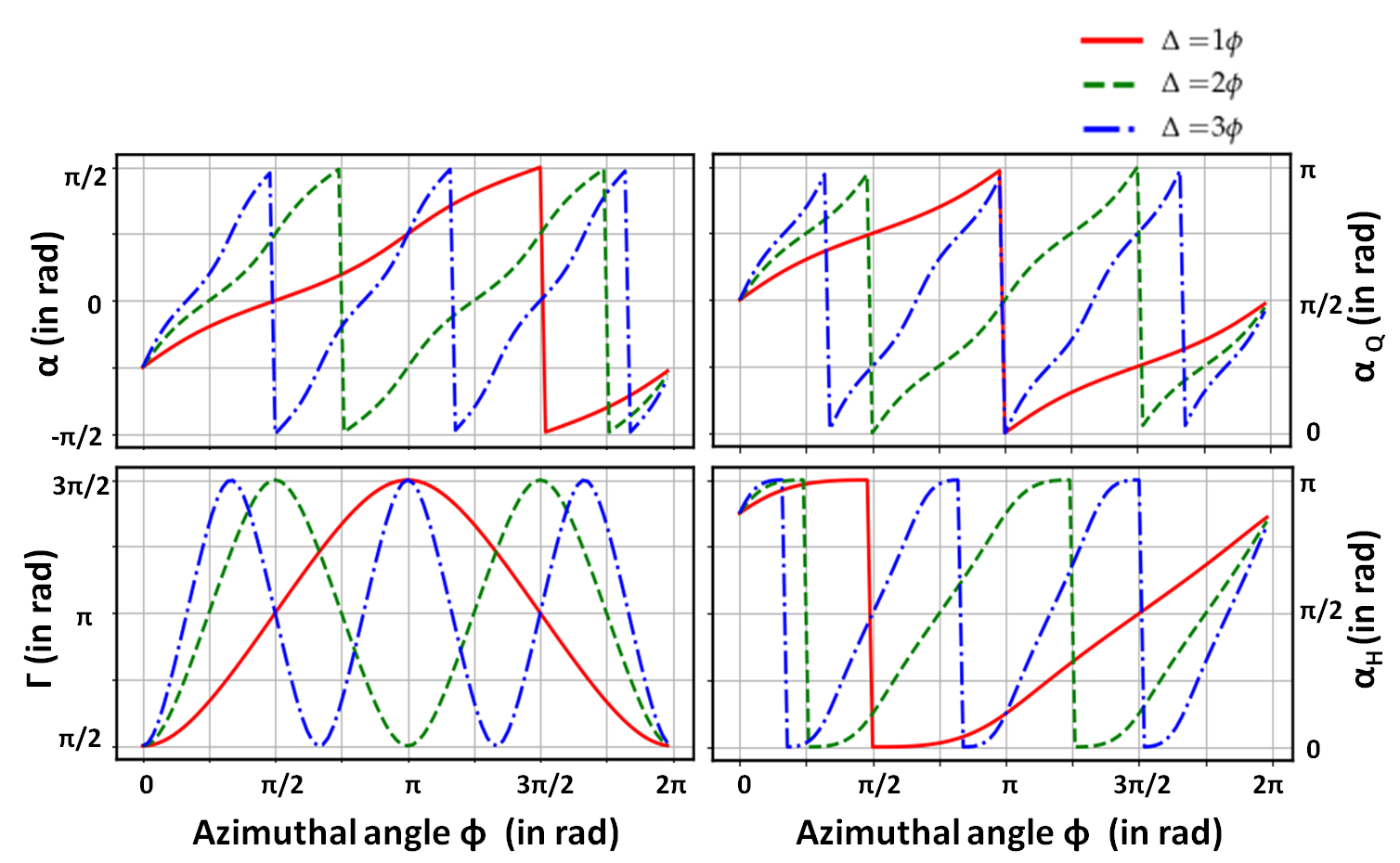}\caption{Azimuthal variation in the retardance$\left(\Gamma\right)$ and fast-axis
orientation$\left(\alpha\right)$ of d-plates $\mathcal{W}_{\Gamma}\left(\alpha\right)$
of Eqs. (\ref{eq:Our_J_Plate_Gamma} and \ref{eq:Our_J_Plate_Alpha}),
for $\frac{m-n}{2}=1,2,3$ is plotted in the left column. For these
d-plates, the corresponding azimuthal variation of fast axes orientation
of QW-s-plates$\left(\alpha_{Q}\left(\phi\right)\right)$ and HW-s-plates
$\left(\alpha_{H}\left(\phi\right)\right)$ required for realizing
them through QHQ arrangement are shown in the right column. \label{fig:Our_J_plate_Variations}}
\end{figure}

Here we demonstrate the construction of such a J-plate for transforming
the SoP $\left|\frac{\pi}{4},0\right\rangle $ to its enantiogyre,
$\left|\frac{3\pi}{4},0\right\rangle $. The required retardance and
fast-axis orientations of Eqs. (\ref{eq:J_Plate_Gamma} and \ref{eq:J_Plate_Alpha})
in this case simplify to:

\begin{align}
{\textstyle \cos\frac{\Gamma}{2}=} & {\textstyle \frac{\cos\Delta}{\sqrt{2}}}\label{eq:Our_J_Plate_Gamma}\\
\tan2\alpha= & {\textstyle -\frac{\cot\Delta}{\sqrt{2}}}\label{eq:Our_J_Plate_Alpha}
\end{align}
where $\Delta=\frac{m-n}{2}\phi$. Figure (\ref{fig:Our_J_plate_Variations})
shows the azimuthal variation of $\Gamma$ and $\alpha$ of the above
equations, for $\Delta=\phi,\,2\phi$ and $3\phi$. The necessary
fast axes variation of the QW-s-plates and HW-s-plates for realizing
these using QHQ arrangement are also shown. The smooth azimuthal variation
of $\alpha_{Q}$ and $\alpha_{H}$ of the s-plates indicate the ease
of their fabrication. The jumps observed in the fast-axis variations
are always of magnitude $\pi$ and arise due to numerical inversion
of Eq. (\ref{eq:Our_J_Plate_Alpha}). Since orientation of fast-axis
is modulo $\pi$, these jumps are experimentally inconsequential.

\begin{figure}[H]
\begin{centering}
\includegraphics[scale=0.35]{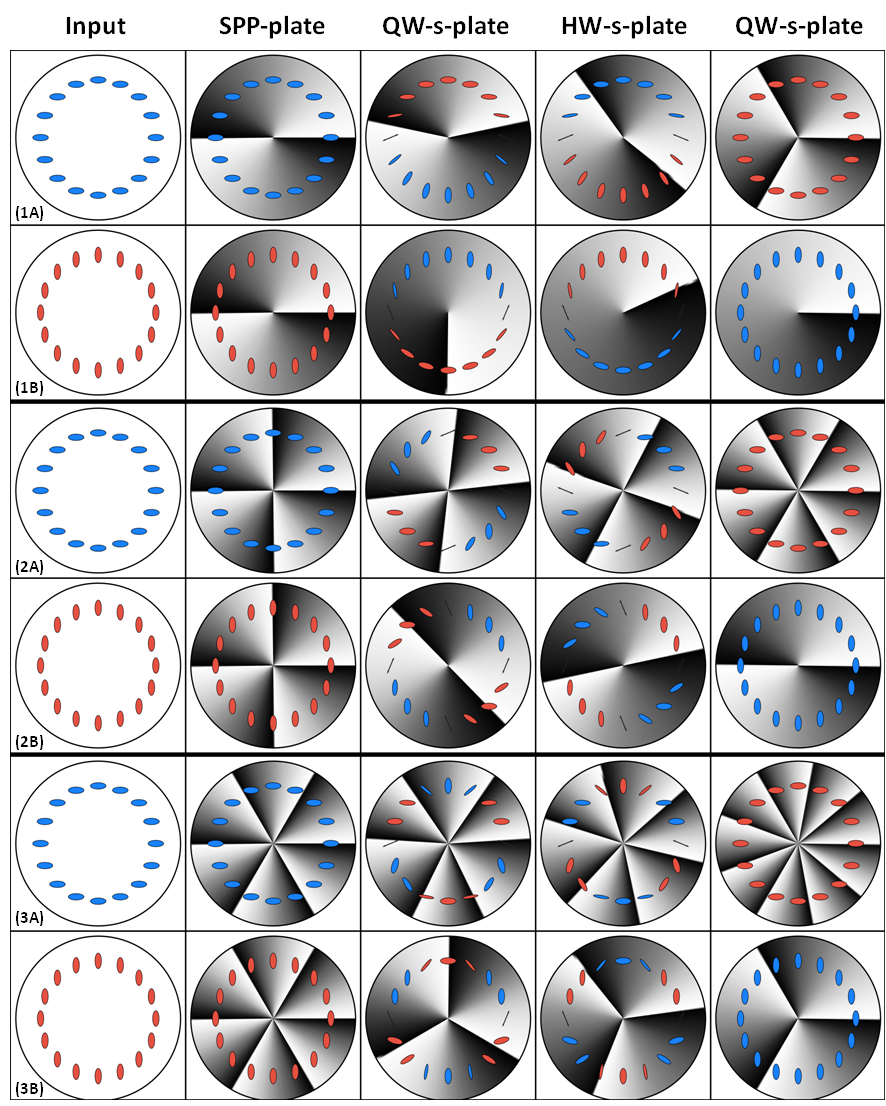} 
\par\end{centering}
\caption{Polarization and phase profile of the light beam in its transverse
plane, after emerging from the spiral phase plate and the QHQ arrangement
of s-plates, set for realizing three different J-plates. The chosen
polarization parameters for the J-plate are $\theta=\frac{\pi}{4}$
and $\varphi=0$, and $\left(m,n\right)$ being $\left(3,1\right)$,
$\left(6,2\right)$ and $\left(9,3\right)$. Rows A corresponds to
the SoP $\vert{\textstyle \frac{\pi}{4},0}\rangle$ and rows B correspond
to its orthogonal SoP $\vert{\textstyle \frac{3\pi}{4},0}\rangle$
as inputs. \label{fig:SoP_Phase_Depiction}}
\end{figure}

We now demonstrate the functioning of the QHQ arrangement towards
realizing the J-plates. Figure (\ref{fig:SoP_Phase_Depiction}) traces
the simulated phase and polarization profile in the transverse plane
of the light beam as it exits through each of the four plates (spiral
phase plate and QHQ-s-plates). The input SoPs considered are $\left|\frac{\pi}{4},0\right\rangle $
and its orthogonal state$\left|\frac{3\pi}{4},\pi\right\rangle $,
shown as rows (A) and (B) respectively. Rows (1), (2) and (3) correspond
to the J-plates having $\left(m,n\right)=$$\left(3,1\right)$,$\left(6,2\right)$
and $\left(9,3\right)$ respectively. These J-plates correspond to
the d-plates of Fig. (\ref{fig:Our_J_plate_Variations}) together
with the spiral phase plate of order $2$, $4$ and $6$ respectively.
In these figures, the polarization profile along the radial direction
is uniform and hence is suppressed for visual clarity. The greyscale
color coding depicts the phase in the transverse plane. In each of
the six cases, even though the input scalar beam gets converted to
vector beam in transit, the beam at the exit plane of the final plate
is once again scalar but with the intended phase distribution. This
demonstrates the utility of the QHQ-s-plates towards mimicking the
J-plate.

To summarize this subsection, a generalized version of q-plates, for
affecting polarization-dependent phase profile to the input light
beam is mathematically elucidated through Eqs. (\ref{eq:U_Transformation}
and \ref{eq:U_Transformation_Antipodal}). Parameters of the d-plate
that assists this transformation are deduced in Eqs. (\ref{eq:J_Plate_Gamma}
and \ref{eq:J_Plate_Alpha}). A simple, albeit non-trivial illustration
of this idea has been demonstrated numerically in realizing the ``J-plate''
proposed recently in ref. \citep{devlin2017arbitrary}.

\section{\label{sec:Summary-and-Conclusions}Summary and Conclusions}

Optical elements which affect the polarization and phase of the light
beam, without affecting its intensity belong to the class of ``waveplates''.
They are characterized by their retardance and orientation of fast-axis,
and have been classified here into different kinds based on the spatial
variation of these parameters. Waveplates exhibiting inhomogeneity
in both their retardance and fast-axis orientation, referred to as
d-plates here, have proved to be of immense utility in the recent
years. The interplay of phase and polarization of the light, as it
traverses through the d-plate, can be gainfully exploited towards
realizing exotic structured light beams.

Fabrication of these d-plates, however, continues to be a formidable
challenge, in contrast to singly inhomogeneous waveplates (s-plates),
whose manufacture has been standardized in multiple ways. In this
article a novel method for realizing ``effective'' d-plate involving
only s-plates, has been proposed. This is achieved by establishing
the equivalence of a d-plate and a gadget involving a HW-s-plate placed
between two identically oriented QW-s-plates, using SU(2) formalism.
Given a d-plate, -with an arbitrary spatial variation in both retardance
and fast-axis orientation,- the necessary HW-s-plate and QW-s-plates
for realizing it have been identified. An important advantage of this
approach is that, using the same set of three s-plates, a variety
of inequivalent d-plates can be realized by mere change of their relative
orientations. The methodology is generic enough to be employed with
any of the s-plate fabrication techniques. Further, the ability to
dynamically tune the retardance of s-plates (for eg. with voltage
controlled liquid-crystal based s-plates) allows for QHQ-s-plates
to realize d-plates for different wavelengths. 

The versatility and potential of this method has been exhibited in
two steps: firstly, specific d-plates required for tailoring individual
aspects of light have been identified. For each of these d-plates,
the required s-plates have been identified and their equivalence with
the d-plates has been analytically and numerically established.

Another significant contribution of this article has been in fashioning
d-plates for (i) realizing an arbitrary complex field amplitude distribution
and (ii) polarization-dependent phase manipulation.

In the first application, starting from a known electric field distribution,
a generic polarimetric approach for realizing arbitrary complex electric
field distribution has been presented. As a concrete example, LG beams
and HG beams of higher order have been sculpted from the horizontally
polarized fundamental Gaussian beam, using a specially designed d-plate
and a polarizer. The ratio of input and output beam waists is seen
to play an important role in deciding the conversion efficiency and
needs to be optimized.

The second application presented here generalizes the functioning
of q-plates. The phase-polarization interplay seen in waveplates has
been exploited in designing a d-plate that functions like a polarization-dependent
spatially varying phase plate. This d-plate imparts distinct phase
distribution on the input light beam depending on its polarization.

The d-plates discussed here, although of significant interest in themselves,
are not exhaustive but merely cursory pointing to their rich class.
QHQ-s-plates offer an easy method of realizing any such d-plates,
and we believe they will greatly advance the experimental state of
the art in structuring of light and application involving structured
light beams.

\appendix

\section{\label{sec:Single-effective-waveplate.}Single effective waveplate.}

The Jones matrices of HW-plate (retardance $\Gamma=\pi$) is given
by (see Eq.(\ref{eq:Waveplate_SU2}))

$\mathcal{W}_{\pi}\left(\alpha\right)=i\left(\sigma_{x}\cos2\alpha+\sigma_{y}\sin2\alpha\right)$

and the Jones matrix of QW-plate (retardance $\Gamma=\frac{\pi}{2}$)
is given by:

$\mathcal{W}_{\frac{\pi}{2}}\left(\alpha\right)=\frac{1}{\sqrt{2}}\left(\bm{I}+i\left(\sigma_{x}\cos2\alpha+\sigma_{y}\sin2\alpha\right)\right)$

In QHQ arrangement, an HW-plate oriented at any angle say $\alpha_{H}$
is sandwiched between two identically oriented QW-plates $\alpha_{Q}$.

\begin{equation}
\mathcal{W}=\mathcal{W}_{\frac{\pi}{2}}\left(\alpha_{Q}\right)\cdot\mathcal{W}_{\pi}\left(\alpha_{H}\right)\cdot\mathcal{W}_{\frac{\pi}{2}}\left(\alpha_{Q}\right)
\end{equation}

The resultant matrix $\mathcal{W}$ of QHQ arrangement can be evaluated
by making use of the Pauli-matrices identity namely: $\sigma_{j}\cdot\sigma_{k}=\delta_{jk}I+i\epsilon_{jkl}\sigma_{l}$,
given by:

\begin{widetext}

\begin{equation}
\mathcal{W}=\left[\begin{array}{cc}
-\cos\left(2\alpha_{Q}-2\alpha_{H}\right)-i\sin\left(2\alpha_{Q}-2\alpha_{H}\right)\sin2\alpha_{Q} & i\sin\left(2\alpha_{Q}-2\alpha_{H}\right)\cos2\alpha_{Q}\\
i\sin\left(2\alpha_{Q}-2\alpha_{H}\right)\cos2\alpha_{Q} & -\cos\left(2\alpha_{Q}-2\alpha_{H}\right)+i\sin\left(2\alpha_{Q}-2\alpha_{H}\right)\sin2\alpha_{Q}
\end{array}\right]
\end{equation}

\end{widetext}

The resultant matrix of QHQ arrangement, eq. (), is symmetric and
has purely imaginary off-diagonal elements, similar to the Eq. (\ref{eq:SU2_Action}).
This arrangement is therefore equivalent to a single waveplate $\mathcal{W}_{\Gamma_{e}}\left(\alpha_{e}\right)$,
whose effective retardance $\Gamma_{e}$ and an effective fast-axis
orientation $\alpha_{e}$ can be derived from Eqs. (\ref{eq:Gamma_E}
and \ref{eq:Alpha_E}), and they are:

\begin{align}
\Gamma_{e} & =2\pi+4\left(\alpha_{Q}-\alpha_{H}\right)\\
\alpha_{e} & =\alpha_{Q}+\frac{\pi}{4}
\end{align}

\section{\label{sec:Rotating-the-waveplate}Rotating the waveplate}

Consider a d-plate whose spatial variation in fast-axis orientation
and retardance are respectively given by $\alpha\left(r,\phi\right)$
and $\Gamma\left(r,\phi\right)$. Mathematically, given the fast-axis
orientation $\alpha$, the fast-axis in the X-Y plane is given by
$\boldsymbol{\hat{k}}\left(\alpha\right)=\cos\alpha\boldsymbol{\hat{e}_{x}}+\sin\alpha\boldsymbol{\hat{e}_{y}}$,
where $\boldsymbol{\hat{e}_{x}}$ and $\boldsymbol{\hat{e}_{y}}$
are unit vectors along the $X$ and $Y$ directions: 
\begin{equation}
\boldsymbol{\hat{k}}\left(r,\phi\right)=\cos\alpha\left(r,\phi\right)\boldsymbol{\hat{e}_{x}}+\sin\alpha\left(r,\phi\right)\boldsymbol{\hat{e}_{y}}=\left[\begin{array}{c}
\cos\alpha\left(r,\phi\right)\\
\sin\alpha\left(r,\phi\right)
\end{array}\right]
\end{equation}
Now, consider rotating this plate by an angle, say $\delta$. Because
of this rotation, the field of vectors changes to $\boldsymbol{\hat{k}}\left(r,\phi-\delta\right)$.
The rotated vectors in the old coordinates is given by

\begin{equation}
\boldsymbol{\hat{k}}_{new}\left(r,\phi\right)=\mathcal{R}\left(\delta\right)\cdot\boldsymbol{\hat{k}}\left(r,\phi-\delta\right)
\end{equation}
where $\mathcal{R}\left(\delta\right)=\left[\begin{array}{cc}
\cos\delta & \sin\delta\\
-\sin\delta & \cos\delta
\end{array}\right]$ is the rotation matrix about the $Z$-direction. Upon simplification,
we have,

\begin{equation}
\boldsymbol{\hat{k}}_{new}\left(r,\phi\right)=\left[\begin{array}{c}
\cos\left(\alpha\left(r,\phi-\delta\right)+\delta\right)\\
\sin\left(\alpha\left(r,\phi-\delta\right)+\delta\right)
\end{array}\right]
\end{equation}
From which the fast-axis orientation $\alpha_{new}\left(r,\phi\right)$
after the rotation can be obtained as:

\begin{equation}
\alpha^{new}\left(r,\phi\right)=\alpha\left(r,\phi-\delta\right)+\delta\label{eq:Rotated_Fast_Axis_Orientation}
\end{equation}

Retardance being a scalar, transforms simply as.

\begin{equation}
\Gamma^{new}\left(r,\phi\right)=\Gamma\left(r,\phi-\delta\right)\label{eq:Rotated_Retardance}
\end{equation}

\section{Determining the Jones matrix\label{ap:J_plate}}

The Jones matrix of the unitary transformation in the $\left\{ \vert H\rangle,\vert V\rangle\right\} $
basis is given by:

\begin{equation}
\mathcal{U}_{\left\{ \vert H\rangle,\vert V\rangle\right\} }=\left[\begin{array}{cc}
\langle H\vert\mathcal{U}\vert H\rangle & \langle H\vert\mathcal{U}\vert V\rangle\\
\langle H\vert\mathcal{U}\vert H\rangle & \langle V\vert\mathcal{U}\vert V\rangle
\end{array}\right]
\end{equation}

Given the unitary transformation $\mathcal{U}$ of Eqs. (\ref{eq:U_Transformation}
and \ref{eq:U_Transformation_Antipodal}), its action on $\vert H\rangle,\vert V\rangle$
can be identified by expressing them in the $\left\{ \vert\theta,\varphi\rangle,\vert\pi-\theta,\pi+\varphi\rangle\right\} $
basis:

\begin{align}
\vert H\rangle & =a\vert\theta,\varphi\rangle+b\vert\pi-\theta,\pi+\varphi\rangle\\
\vert V\rangle & =c\vert\theta,\varphi\rangle+d\vert\pi-\theta,\pi+\varphi\rangle
\end{align}

\begin{eqnarray*}
a & = & \frac{1}{\sqrt{2}}{\textstyle \left(\sin\frac{\theta}{2}e^{-i\frac{\varphi}{2}}+\cos\frac{\theta}{2}e^{i\frac{\varphi}{2}}\right)},\\
b & = & -\frac{i}{\sqrt{2}}{\textstyle \left(\cos\frac{\theta}{2}e^{-i\frac{\varphi}{2}}-\sin\frac{\theta}{2}e^{i\frac{\varphi}{2}}\right),}\\
c & = & \frac{i}{\sqrt{2}}{\textstyle \left(\sin\frac{\theta}{2}e^{-i\frac{\varphi}{2}}-\cos\frac{\theta}{2}e^{i\frac{\varphi}{2}}\right),}\\
d & = & \frac{1}{\sqrt{2}}{\textstyle \left(\cos\frac{\theta}{2}e^{-i\frac{\varphi}{2}}+\sin\frac{\theta}{2}e^{i\frac{\varphi}{2}}\right).}
\end{eqnarray*}

Expressing $\vert\pi-\theta,\varphi\rangle$ and $\vert\theta,\pi-\varphi\rangle$
in the $\vert H\rangle,\vert V\rangle$ basis:

\begin{eqnarray*}
\vert\pi-\theta,\varphi\rangle & = & a\vert H\rangle+c\vert V\rangle\\
\vert\theta,\pi-\varphi\rangle & = & b\vert H\rangle+d\vert V\rangle
\end{eqnarray*}

\begin{equation}
\mathcal{U}_{\left\{ \vert H\rangle,\vert V\rangle\right\} }=\left[\begin{array}{cc}
e^{i\psi}a^{2}+e^{i\chi}b^{2} & \quad e^{i\psi}ac+e^{i\chi}bd\\
e^{i\psi}ac+e^{i\chi}bd & \quad e^{i\psi}c^{2}+e^{i\chi}d^{2}
\end{array}\right]
\end{equation}

From this the determinant of the $\mathcal{U}_{\left\{ \vert H\rangle,\vert V\rangle\right\} }$matrix
is $e^{i\left(\psi+\chi\right)}$ and this unitary matrix can be converted
into a SU(2) matrix $\mathcal{W}_{\Gamma}\left(\alpha\right)$ by
multiplying $\mathcal{U}_{\left\{ \vert H\rangle,\vert V\rangle\right\} }$
with phase element as:

\begin{eqnarray*}
\mathcal{W}_{\Gamma}\left(\alpha\right) & = & e^{-i\frac{\left(\psi+\chi\right)}{2}}\cdot\mathcal{U}_{\left\{ \vert H\rangle,\vert V\rangle\right\} }\\
 & = & e^{-i\frac{\left(\psi+\chi\right)}{2}}\left[\begin{array}{cc}
e^{i\psi}a^{2}+e^{i\chi}b^{2} & \quad e^{i\psi}ac+e^{i\chi}bd\\
e^{i\psi}ac+e^{i\chi}bd & \quad e^{i\psi}c^{2}+e^{i\chi}d^{2}
\end{array}\right]
\end{eqnarray*}

From this, the required retardance $\Gamma\left(r,\phi\right)$ and
fast-axis $\alpha\left(r,\phi\right)$ of this d-plate can be extracted
through Eqs. (\ref{eq:Gamma_E} and \ref{eq:Alpha_E}).


\begin{thebibliography}{10}

\bibitem{pancharatnam1956generalized}
Shivaramakrishnan Pancharatnam.
\newblock Generalized theory of interference and its applications.
\newblock In {\em Proceedings of the Indian Academy of Sciences-Section A},
  volume~44, pages 398--417. Springer, 1956.

\bibitem{berry1987adiabatic}
Michael~V Berry.
\newblock The adiabatic phase and pancharatnam's phase for polarized light.
\newblock {\em Journal of Modern Optics}, 34(11):1401--1407, 1987.

\bibitem{zhan2009cylindrical}
Qiwen Zhan.
\newblock Cylindrical vector beams: from mathematical concepts to applications.
\newblock {\em Advances in Optics and Photonics}, 1(1):1--57, 2009.

\bibitem{qiwen2013vectorial}
Zhan Qiwen.
\newblock {\em Vectorial optical fields: Fundamentals and applications}.
\newblock World scientific, 2013.

\bibitem{marrucci2006optical}
Lorenzo Marrucci, C~Manzo, and D~Paparo.
\newblock Optical spin-to-orbital angular momentum conversion in inhomogeneous
  anisotropic media.
\newblock {\em Physical review letters}, 96(16):163905, 2006.

\bibitem{andrews2011structured}
David~L Andrews.
\newblock {\em Structured light and its applications: An introduction to
  phase-structured beams and nanoscale optical forces}.
\newblock Academic press, 2011.

\bibitem{rubinsztein2016roadmap}
Halina Rubinsztein-Dunlop, Andrew Forbes, Michael~V Berry, Mark~R Dennis,
  David~L Andrews, Masud Mansuripur, Cornelia Denz, Christina Alpmann, Peter
  Banzer, Thomas Bauer, et~al.
\newblock Roadmap on structured light.
\newblock {\em Journal of Optics}, 19(1):013001, 2016.

\bibitem{rosales2018review}
Carmelo Rosales-Guzm{\'a}n, Bienvenu Ndagano, and Andrew Forbes.
\newblock A review of complex vector light fields and their applications.
\newblock {\em Journal of Optics}, 20(12):123001, 2018.

\bibitem{dorn2003sharper}
Ralf Dorn, Susanne Quabis, and Gerd Leuchs.
\newblock Sharper focus for a radially polarized light beam.
\newblock {\em Physical review letters}, 91(23):233901, 2003.

\bibitem{allen1992orbital}
Les Allen, Marco~W Beijersbergen, RJC Spreeuw, and JP~Woerdman.
\newblock Orbital angular momentum of light and the transformation of
  laguerre-gaussian laser modes.
\newblock {\em Physical Review A}, 45(11):8185, 1992.

\bibitem{shen2019optical}
Yijie Shen, Xuejiao Wang, Zhenwei Xie, Changjun Min, Xing Fu, Qiang Liu, Mali
  Gong, and Xiaocong Yuan.
\newblock Optical vortices 30 years on: Oam manipulation from topological
  charge to multiple singularities.
\newblock {\em Light: Science \& Applications}, 8(1):1--29, 2019.

\bibitem{bliokh2015spin}
K~Yu Bliokh, FJ~Rodr{\'\i}guez-Fortu{\~n}o, Franco Nori, and Anatoly~V Zayats.
\newblock Spin--orbit interactions of light.
\newblock {\em Nature Photonics}, 9(12):796, 2015.

\bibitem{liu2017photonic}
Yachao Liu, Yougang Ke, Hailu Luo, and Shuangchun Wen.
\newblock Photonic spin hall effect in metasurfaces: a brief review.
\newblock {\em Nanophotonics}, 6(1):51--70, 2017.

\bibitem{shitrit2013spin}
Nir Shitrit, Igor Yulevich, Elhanan Maguid, Dror Ozeri, Dekel Veksler, Vladimir
  Kleiner, and Erez Hasman.
\newblock Spin-optical metamaterial route to spin-controlled photonics.
\newblock {\em Science}, 340(6133):724--726, 2013.

\bibitem{aiello2015transverse}
Andrea Aiello, Peter Banzer, Martin Neugebauer, and Gerd Leuchs.
\newblock From transverse angular momentum to photonic wheels.
\newblock {\em Nature Photonics}, 9(12):789--795, 2015.

\bibitem{kozawa2010optical}
Yuichi Kozawa and Shunichi Sato.
\newblock Optical trapping of micrometer-sized dielectric particles by
  cylindrical vector beams.
\newblock {\em Optics Express}, 18(10):10828--10833, 2010.

\bibitem{roxworthy2010optical}
Brian~J Roxworthy and Kimani~C Toussaint~Jr.
\newblock Optical trapping with $\pi$-phase cylindrical vector beams.
\newblock {\em New Journal of Physics}, 12(7):073012, 2010.

\bibitem{taylor2015enhanced}
Michael~A Taylor, Muhammad Waleed, Alexander~B Stilgoe, Halina
  Rubinsztein-Dunlop, and Warwick~P Bowen.
\newblock Enhanced optical trapping via structured scattering.
\newblock {\em Nature Photonics}, 9(10):669--673, 2015.

\bibitem{meier2007material}
Matthias Meier, Valerio Romano, and Thomas Feurer.
\newblock Material processing with pulsed radially and azimuthally polarized
  laser radiation.
\newblock {\em Applied Physics A}, 86(3):329--334, 2007.

\bibitem{erhard2018twisted}
Manuel Erhard, Robert Fickler, Mario Krenn, and Anton Zeilinger.
\newblock Twisted photons: new quantum perspectives in high dimensions.
\newblock {\em Light: Science \& Applications}, 7(3):17146, 2018.

\bibitem{Chigrinov2008}
V.G. Chigrinov, V.M. Kozenkov, and H.S. Kwok.
\newblock {\em Photoalignment of Liquid Crystalline Materials: Physics and
  Applications}.
\newblock Wiley Series in Display Technology. Wiley, 2008.

\bibitem{kim2015fabrication}
Jihwan Kim, Yanming Li, Matthew~N Miskiewicz, Chulwoo Oh, Michael~W Kudenov,
  and Michael~J Escuti.
\newblock Fabrication of ideal geometric-phase holograms with arbitrary
  wavefronts.
\newblock {\em Optica}, 2(11):958--964, 2015.

\bibitem{ji2016meta}
Wei Ji, Chun-Hong Lee, Peng Chen, Wei Hu, Yang Ming, Lijian Zhang, Tsung-Hsien
  Lin, Vladimir Chigrinov, and Yan-Qing Lu.
\newblock Meta-q-plate for complex beam shaping.
\newblock {\em Scientific reports}, 6:25528, 2016.

\bibitem{lin2014dielectric}
Dianmin Lin, Pengyu Fan, Erez Hasman, and Mark~L Brongersma.
\newblock Dielectric gradient metasurface optical elements.
\newblock {\em science}, 345(6194):298--302, 2014.

\bibitem{khorasaninejad2016metalenses}
Mohammadreza Khorasaninejad, Wei~Ting Chen, Robert~C Devlin, Jaewon Oh,
  Alexander~Y Zhu, and Federico Capasso.
\newblock Metalenses at visible wavelengths: Diffraction-limited focusing and
  subwavelength resolution imaging.
\newblock {\em Science}, 352(6290):1190--1194, 2016.

\bibitem{scheuer2020optical}
Jacob Scheuer.
\newblock Optical metasurfaces are coming of age: Short-and long-term
  opportunities for commercial applications.
\newblock {\em ACS Photonics}, 2020.

\bibitem{ling2015giant}
Xiaohui Ling, Xinxing Zhou, Xunong Yi, Weixing Shu, Yachao Liu, Shizhen Chen,
  Hailu Luo, Shuangchun Wen, and Dianyuan Fan.
\newblock Giant photonic spin hall effect in momentum space in a structured
  metamaterial with spatially varying birefringence.
\newblock {\em Light: Science \& Applications}, 4(5):e290--e290, 2015.

\bibitem{pal2016tunable}
Mandira Pal, Chitram Banerjee, Shubham Chandel, Ankan Bag, Shovan~K Majumder,
  and Nirmalya Ghosh.
\newblock Tunable spin dependent beam shift by simultaneously tailoring
  geometric and dynamical phases of light in inhomogeneous anisotropic medium.
\newblock {\em Scientific reports}, 6:39582, 2016.

\bibitem{devlin2017arbitrary}
Robert~C Devlin, Antonio Ambrosio, Noah~A Rubin, JP~Balthasar Mueller, and
  Federico Capasso.
\newblock Arbitrary spin-to--orbital angular momentum conversion of light.
\newblock {\em Science}, 358(6365):896--901, 2017.

\bibitem{rafayelyan2017laguerre}
Mushegh Rafayelyan and Etienne Brasselet.
\newblock Laguerre--gaussian modal q-plates.
\newblock {\em Optics letters}, 42(10):1966--1969, 2017.

\bibitem{rubin2019matrix}
Noah~A Rubin, Gabriele D'Aversa, Paul Chevalier, Zhujun Shi, Wei~Ting Chen, and
  Federico Capasso.
\newblock Matrix fourier optics enables a compact full-stokes polarization
  camera.
\newblock {\em Science}, 365(6448):eaax1839, 2019.

\bibitem{arbabi2015dielectric}
Amir Arbabi, Yu~Horie, Mahmood Bagheri, and Andrei Faraon.
\newblock Dielectric metasurfaces for complete control of phase and
  polarization with subwavelength spatial resolution and high transmission.
\newblock {\em Nature nanotechnology}, 10(11):937--943, 2015.

\bibitem{mueller2017metasurface}
JP~Balthasar Mueller, Noah~A Rubin, Robert~C Devlin, Benedikt Groever, and
  Federico Capasso.
\newblock Metasurface polarization optics: independent phase control of
  arbitrary orthogonal states of polarization.
\newblock {\em Physical review letters}, 118(11):113901, 2017.

\bibitem{piccirillo2010photon}
Bruno Piccirillo, Vincenzo D~Ambrosio, Sergei Slussarenko, Lorenzo Marrucci,
  and Enrico Santamato.
\newblock Photon spin-to-orbital angular momentum conversion via an
  electrically tunable q-plate.
\newblock {\em Applied Physics Letters}, 97(24):241104, 2010.

\bibitem{Slussarenko:11}
Sergei Slussarenko, Anatoli Murauski, Tao Du, Vladimir Chigrinov, Lorenzo
  Marrucci, and Enrico Santamato.
\newblock Tunable liquid crystal q-plates with arbitrary topological charge.
\newblock {\em Opt. Express}, 19(5):4085--4090, Feb 2011.

\bibitem{Karimi2009}
Ebrahim Karimi, Bruno Piccirillo, Eleonora Nagali, Lorenzo Marrucci, and Enrico
  Santamato.
\newblock Efficient generation and sorting of orbital angular momentum
  eigenmodes of light by thermally tuned q-plates.
\newblock {\em Applied Physics Letters}, 94(23):231124, 2009.

\bibitem{pancharatnam_1_1955achromatic}
Shivaramakrishnan Pancharatnam.
\newblock Achromatic combinations of birefringent plates part-i.
\newblock In {\em Proceedings of the Indian Academy of Sciences-Section A},
  volume~41, pages 130--136. Springer, 1955.

\bibitem{pancharatnam2_1955achromatic}
Shivaramakrishnan Pancharatnam.
\newblock Achromatic combinations of birefringent plates part-ii.
\newblock In {\em Proceedings of the Indian Academy of Sciences-Section A},
  volume~41, pages 137--144. Springer, 1955.

\bibitem{thorlabs_Pancharatnam_achromat}
Thorlabs super achromatic waveplates.

\bibitem{yi2015addition}
Xunong Yi, Ying Li, Xiaohui Ling, Yachao Liu, Yougang Ke, and Dianyuan Fan.
\newblock Addition and subtraction operation of optical orbital angular
  momentum with dielectric metasurfaces.
\newblock {\em Optics Communications}, 356:456--462, 2015.

\bibitem{delaney2017arithmetic}
Sam Delaney, Mar{\'\i}a~M S{\'a}nchez-L{\'o}pez, Ignacio Moreno, and Jeffrey~A
  Davis.
\newblock Arithmetic with q-plates.
\newblock {\em Applied optics}, 56(3):596--600, 2017.

\bibitem{radhakrishna2019wavelength}
B~Radhakrishna, Gururaj Kadiri, and G~Raghavan.
\newblock Wavelength-adaptable effective q-plates with passively tunable
  retardance.
\newblock {\em Scientific reports}, 9(1):1--9, 2019.

\bibitem{korotkova2013random}
Olga Korotkova.
\newblock {\em Random light beams: theory and applications}.
\newblock CRC press, 2013.

\bibitem{galvez2017complex}
Enrique~J Galvez.
\newblock Complex light beams.
\newblock {\em Deep Imaging in Tissue and Biomedical Materials: Using Linear
  and Nonlinear Optical Methods}, page~31, 2017.

\bibitem{bouchal2003nondiffracting}
Zden{\v{e}}k Bouchal.
\newblock Nondiffracting optical beams: physical properties, experiments, and
  applications.
\newblock {\em Czechoslovak journal of physics}, 53(7):537--578, 2003.

\bibitem{mazilu2010light}
Michael Mazilu, D~James Stevenson, Frank Gunn-Moore, and Kishan Dholakia.
\newblock Light beats the spread:'non-diffracting' beams.
\newblock {\em Laser \& Photonics Reviews}, 4(4):529--547, 2010.

\bibitem{efremidis2019airy}
Nikolaos~K Efremidis, Zhigang Chen, Mordechai Segev, and Demetrios~N
  Christodoulides.
\newblock Airy beams and accelerating waves: an overview of recent advances.
\newblock {\em Optica}, 6(5):686--701, 2019.

\bibitem{granata2010higher}
Massimo Granata, Christelle Buy, Robert Ward, and Matteo Barsuglia.
\newblock Higher-order laguerre-gauss mode generation and interferometry for
  gravitational wave detectors.
\newblock {\em Physical review letters}, 105(23):231102, 2010.

\bibitem{tao2020higher}
Liu Tao, Anna Green, and Paul Fulda.
\newblock Higher-order hermite-gauss modes as a robust flat beam in
  interferometric gravitational wave detectors.
\newblock {\em arXiv preprint arXiv:2010.04338}, 2020.

\bibitem{hell1994breaking}
Stefan~W Hell and Jan Wichmann.
\newblock Breaking the diffraction resolution limit by stimulated emission:
  stimulated-emission-depletion fluorescence microscopy.
\newblock {\em Optics letters}, 19(11):780--782, 1994.

\bibitem{yu2016super}
Wentao Yu, Ziheng Ji, Dashan Dong, Xusan Yang, Yunfeng Xiao, Qihuang Gong, Peng
  Xi, and Kebin Shi.
\newblock Super-resolution deep imaging with hollow bessel beam sted
  microscopy.
\newblock {\em Laser \& Photonics Reviews}, 10(1):147--152, 2016.

\bibitem{simpson1996optical}
NB~Simpson, L~Allen, and MJ~Padgett.
\newblock Optical tweezers and optical spanners with laguerre--gaussian modes.
\newblock {\em Journal of modern optics}, 43(12):2485--2491, 1996.

\bibitem{simpson1997mechanical}
NB~Simpson, K~Dholakia, L~Allen, and MJ~Padgett.
\newblock Mechanical equivalence of spin and orbital angular momentum of light:
  an optical spanner.
\newblock {\em Optics letters}, 22(1):52--54, 1997.

\bibitem{wright2000toroidal}
EM~Wright, J~Arlt, and K~Dholakia.
\newblock Toroidal optical dipole traps for atomic bose-einstein condensates
  using laguerre-gaussian beams.
\newblock {\em Physical Review A}, 63(1):013608, 2000.

\bibitem{rafayelyan2017laguerre_nanostructure}
Mushegh Rafayelyan, Titas Gertus, and Etienne Brasselet.
\newblock Laguerre-gaussian quasi-modal q-plates from nanostructured glasses.
\newblock {\em Applied Physics Letters}, 110(26):261108, 2017.

\bibitem{damask2004polarization}
Jay~N Damask.
\newblock {\em Polarization optics in telecommunications}, volume 101.
\newblock Springer Science \& Business Media, 2004.

\bibitem{Bettegowda2017}
Radhakrishna Bettegowda.
\newblock Prescription for transforming polarization states of light using
  two-quarter waveplates.
\newblock {\em Optical Engineering}, 56(3):034110, 2017.

\bibitem{salazar2018trajectories}
Karol Salazar-Ariza and Rafael Torres.
\newblock Trajectories on the poincar{\'e} sphere of polarization states of a
  beam passing through a rotating linear retarder.
\newblock {\em JOSA A}, 35(1):65--72, 2018.

\bibitem{simon1989universal}
R~Simon and N~Mukunda.
\newblock Universal su (2) gadget for polarization optics.
\newblock {\em Physics Letters A}, 138(9):474--480, 1989.

\bibitem{bhandari1997polarization}
Rajendra Bhandari.
\newblock Polarization of light and topological phases.
\newblock {\em Physics Reports}, 281(1):1--64, 1997.

\bibitem{beijersbergen1994helical}
MW~Beijersbergen, RPC Coerwinkel, M~Kristensen, and JP~Woerdman.
\newblock Helical-wavefront laser beams produced with a spiral phaseplate.
\newblock {\em Optics Communications}, 112(5-6):321--327, 1994.

\end{thebibliography}
\end{document}